\documentstyle[preprint,aps,epsf,eqsecnum]{revtex}
\def\setepsfscale#1{\def\epsfsize##1##2{#1##1}}
\newcommand{\postscript}[1]
 {\centerline{\epsfbox{#1}}}
\begin{document}
\draft
\preprint{\vbox{
\hbox{IFT-P.060/93}
\hbox{IFUSP/P-1074}
\hbox{hep-ph/9310yyy}
\hbox{September 1993}
}}
\title{
Some consequences in weak processes of three generations \\mixing in
the leptonic sector
}
\author{O.L.G. Peres$^{a}$, V. Pleitez$^{a}$
\\and \\R. Zukanovich Funchal$^{b}$ }
\address{
$^{a}$Instituto de F\'\i sica Te\'orica \\
Universidade Estadual Paulista \\
Rua Pamplona, 145 \\
01405-900--S\~ao Paulo, SP \\
Brazil \\
$^{b}$ Instituto de F\'\i sica da Universidade de S\~ao Paulo\\
01498-970 C.P. 20516--S\~ao Paulo, SP\\
Brazil\\
}
\maketitle
\begin{abstract}
We investigate the sensitivity of some weak processes to a
Cabibbo--Kobayashi--Maskawa mixing in the leptonic sector. Values for
mixing angles and masses compatible with several experimental accelerator data
were found. We discuss in this context neutrino oscillations, cosmological
and astrophysical consequences as well.
\end{abstract}
\pacs{PACS numbers:14.60-z, 12.15.Ff, 13.35+s}
\newpage

\narrowtext

\section{Introduction} 
\label{sec:int}
The physics of the $\tau$ lepton will provide in the near future
evidence concerning the question if this lepton, with its
neutrino partner, is a sequential lepton or not.
Although the existing data indicate a positive answer to that
question, the structure of the weak  interaction of the third lepton
family deserves more  detailed experimental and theoretical studies.
As was stressed in Ref.~\cite{eppz}, all experiments are
internally consistent with the standard model. Notwithstanding, it is
well known that the accuracy of the $\tau$ data is still poor and it
should be possible that new physics will come up when the proposed
$\tau$-Charm factories give new and more accurate data about $\tau$
decays and properties~\cite{p}.

One important issue concerns lepton universality. The $e-\mu$
universality is well established in $\pi$-decays given a ratio~\cite{b2}
\begin{equation}
G_e/G_\mu=0.9985\pm0.0015.
\label{emuniv}
\end{equation}
On the other hand, the current data on $\tau-\mu$ universality have not
the same precision and one still can speculate on a possible departure from
$\tau-\mu$ universality.

A quantity which parametrizes the $\tau-\mu$
universality is
\begin{equation}
\left(\frac{G_\tau}{G_\mu}\right)^2=\left(\frac{\tau_\mu}
{\tau_\tau}\right)
\left(\frac{m_\mu}{m_\tau}\right)^5\frac{B^{\tau e}}{B^{\mu e}},
\label{2}
\end{equation}
where $G_\tau$ and $G_\mu$ are, respectively, the coupling constants
of the $\tau$
and $\mu$ to the charged weak current respectively. We will use the
notation $B^{ab}=B.R.(a\to b+X)$ where $X$ are appropriate particles.
A similar notation is used for the partial width $\Gamma^{ab}$.

The new values for
the $\tau$ mass $m_\tau=1776.9^{+0.4}_{-0.5}\pm 0.2$ MeV~\cite{bes} or
$m_\tau=1771.3\pm2.4\pm1.4$ MeV~\cite{argus} imply that a possible
deviation from
$\tau-\mu$  universality is reduced from 2.5$\sigma$ to 1.7$\sigma$.
Explicitly with the BES data one has
\begin{equation}
\left( \frac{G_\tau}{G_\mu}\right)^2=0.960\pm0.024.
\label{tmu}
\end{equation}
\newpage
Two other experimental quantities which we will use later on
are~\cite{pdg}
\begin{equation}
\frac{B^{\tau\mu}}{B^{\tau e}}=0.98\pm0.02.
\label{tmu2}
\end{equation}
and
\begin{equation}
\frac{B^{\pi e}}{B^{\pi\mu}}=(1.218\pm0.014) \times
10^{-4},
\label{pi}
\end{equation}
A more recent data for the ratio is given in (\ref{pi})
is~\cite{dib1}
\begin{equation}
1.2265\pm0.0034\,(\mbox{stat})\pm0.0044\,(\mbox{sys})\times10^{-4}
\label{recent}
\end{equation}
 however we will use, for consistence, the world average of
Ref.~\cite{pdg} for all quantities.

In this work, we will consider that if neutrinos are massive in the
standard electroweak model~\cite{ws},
a mixing similar to the Cabibbo-Kobayashi-Maskawa (CKM) one
in the quark sector~\cite{km}, may occur with three lepton
generations.
We show in this paper that such a mixing it is not ruled out by current
data and can be used to explain a possible deviation from $\tau-\mu$
universality.

We must nevertheless stress that the analysis we will put forward
in this work is valid even if $\tau-\mu$ universality is confirmed.
Only the numerical values that we give will change.

The introduction of a mixing among three generations makes room for a
third massive neutrino with a mass, as we will see, beyond the current
limit of what
is consider, without mixing, the ``tau neutrino''~\cite{argus,argus2}.

The existence of neutral heavy leptons has been discussed in the
literature since about twenty years ago~\cite{teo}. Usually, the
heavy neutrino is assumed to mix with the electron and muon families
with either $V-A$ or $V+A$ charged and neutral currents. The heavy
neutrino could be a higher generation neutrino, associated with a yet
unobserved heavy charged lepton, a right-handed neutrino,
etc.~\cite{mg1}. Another possibility usually considered, is that in
addition to the three left-handed lepton doublets one includes $k$
neutrinos singlets $\chi_j$, $j=1,...,k$ \cite{mg2,lsl}. This led to
neutrino mixing without affecting the $Z^0$ width.

Here we will suppose that only 3 right-handed singlets have been
added and that there is no Majorana mass terms among the
right-handed singlets. This is the simplest extension of the standard
model which includes massive Dirac neutrinos.
 The only supposition we will make is that
\begin{equation}
0\lower.7ex\hbox{$\;\stackrel{\textstyle<}{\sim}\;$}m_{\nu_1}\approx
m_{\nu_2}\ll m_{\nu_3}.
\label{aprox}
\end{equation}
This will be assumed not only for the sake of simplicity but because
it is an interesting possibility. The two light neutrinos could be
detected in oscillation experiments~\cite{cr}. We will not make any
further assumptions concerning the mixing matrix elements nor the
$m_{\nu_3}$ value but leave them to be constrained by several
experimental data. This means that we will fit in our analysis
, at the same time, two
matrix elements and a mass.

We stress that it is important to specify the model one is
considering when comparing with experimental data. A purely
model independent approach cannot be definitive.

The outline of this work is as follows. In Sec.~\ref{sec:lepmix} we
consider the effect of mixing in the partial
width for the muon decay, leptonic tau decays and pion decays. These
are the quantities which are calculated theoretically. In
Sec.~\ref{sec:ma} we compare our
theoretical results with experimental data (within $1\sigma$)
for $\tau$ and pion decays
deriving from this comparison regions of permitted mixing angles and
mass.  In Sec.~\ref{subsec:z0} we take into account the
constraints coming from the $Z^0$ invisible partial width.
In Sec. \ref{subsec:bound} we
discuss the current limit of ``$m_{\nu_\tau}$'' on the light of
the mixing angles obtained in Sec~\ref{sec:ma}, showing that it is
possible to go beyond the current limit of 31 MeV without
contradicting experimental results. Neutrino oscillations are consider in
Sec.\ref{subsec:osc},
cosmological and astrophysical
constraints are discussed in Sec. \ref{subsec:cosmo} and Sec.
\ref{subsec:astro}
respectively. Finally, the last section is devoted to our conclusions.

\section{Three generations mixing in the leptonic sector}
\label{sec:lepmix}
In this paper we will consider only Dirac neutrinos.
The general effects of the Cabibbo-Kobayashi-Maskawa mixing in the
leptonic sector were considered in Ref.~\cite{kg,ro,ng}. In particular
the  effects of such a mixing for the case of leptonic decays of the
$\mu$ were explicitly considered in Ref.~\cite{pk1}, and in
Ref.~\cite{ssi} for the $\tau$ lepton case.

In fact, it has been shown by Shrock~\cite{ro} that if neutrinos
have nonzero masses the semi-leptonic decays $h'\to h+l+\nu_l$
consist of an incoherent sum of the separate modes $h'\to h+l+\nu_i$,
where $\nu_l=\nu_e,\nu_\mu,\nu_\tau$
are weak eigenstates and $\nu_i,\,i=1,2,3$ are the mass eigenstates
allowed by phase space. These states are related by
$\nu_l=\sum_iV_{li}\nu_i$, $V$ being the unitary leptonic
CKM matrix. Explicitly, this mixing
matrix is defined for the three generations case as
\begin{equation}
\left(\begin{array}{c}
\nu_e \\ \nu_\mu \\ \nu_\tau
\end{array}\right)=\left(
\begin{array}{ccc}
V_{e1}    & V_{e2}    & V_{e3} \\
V_{\mu1} & V_{\mu2}  & V_{\mu3} \\
V_{\tau1}& V_{\tau2} &V_{\tau3}
\end{array}\right)
\left(
\begin{array}{c}\nu_1\\ \nu_2 \\ \nu_3\end{array}\right).
\label{4}
\end{equation}

It is well known that there are several parametrizations of the
CKM matrix. Here we will use the Maiani one
{}~\cite{pdg,maiani}. This form has the feature that
when two of such angles vanish the mixing matrix reduces
to the usual Cabibbo-like mixing matrix of two generations. Using the
usual notation $c_{ij}=\cos\theta_{ij},s_{ij}=\sin\theta_{ij}$, with
$i$ and $j$ being generation labels $i,j=1,2,3$, setting the phase
$\delta_{13}=0$ for the sake of simplicity and making the
substitution $c_{12}\to c_\theta,c_{13}\to c_\beta$ and $c_{23}\to
c_\gamma$ we have
\begin{equation}
\left(
\begin{array}{ccc}
c_\theta c_\beta \; \; \; & s_\theta c_\beta  \; \; & s_\beta \\
-s_\theta c_\gamma-c_\theta s_\gamma s_\beta \; \; \; & \; \; \;
c_\theta c_\gamma
-s_\theta s_\gamma s_\beta\; \; \;  & \; \; \;
s_\gamma c_\beta  \\
s_\theta s_\gamma -c_\theta c_\gamma s_\beta \; \; \; & \; \; \;
-c_\theta s_\gamma
-s_\theta c_\gamma s_\beta\; \; \;  & \; \; \;
c_\gamma c_\beta
\end{array}
\right).
\label{ckm}
\end{equation}

Assuming that only one of the neutrinos,
say $\nu_3$, is sufficiently massive we can write for the decay
probability of a charged lepton $l'$ into the charged lepton $l$ and
neutrinos $\nu_i\bar\nu_j$, the following expression~\cite{ssi},
\begin{eqnarray}
\Gamma(l'\to l\bar\nu_l\nu_{l'})&=&\frac{G^2m^5_{l'}}{192\pi^3}\left\{
\left(\vert V_{l'1}\vert^2+\vert V_{l'2}\vert^2\right)
\left(\vert V_{l1}\vert^2+\vert
V_{l2}\vert^2\right)\,\Gamma^{l'l}_{00}\right.\nonumber \\
&+&
\left.\left(\vert V_{l3}\vert^2[\vert V_{l'1}\vert^2+\vert V_{l'2}\vert^2]
+\vert V_{l'3}\vert^2[\vert V_{l1}\vert^2+\vert V_{l2}\vert^2]
\right)\,\Gamma^{l'l}_{03}\right.\nonumber \\
&+&\left.\vert V_{l'3}\vert^2\vert
V_{l3}\vert^2\,\Gamma^{l'l}_{33}\right\},
\label{decay}
\end{eqnarray}
with $l'=\mu,\tau$ and $l=e,\mu$ for the $\tau$ decay
and $l=e$ for the muon decay. Notice that $G^2$, although it is still
defined as $G^2/\sqrt2=g^2/8m^2_W$ in Eq. (\ref{decay}) is not equal to
the muon decay constant, $G_\mu$. We shall return to this point later.
In (\ref{decay}) we have defined the integrals
\begin{equation}
\Gamma^{l'l}_{00}=2\int^{x_M}_{x_m}(x^2-B)^{\frac{1}{2}}[x(3k-2x)-B]dx
\label{r0}
\end{equation}
\begin{eqnarray}
\Gamma^{l'l}_{03} & = &
2\int^{x_M}_{x_m}(x^2-B)^{\frac{1}{2}}\frac{(k-\delta_{3 l'}^2-x)}{(k-x)^3}
\left[(k-\delta_{3 l'}^2-x)^2x(k-x) \right. \nonumber \\
 & + &\left. [(k-x)^2+\delta_{3 l'}^2(k-x)-2\delta_{3 l'}^4] (2kx-x^2-B)\right]
   dx
\label{r3}
\end{eqnarray}
\begin{eqnarray}
\Gamma^{l'l}_{33} & = &
2\int^{x_M}_{x_m}\frac{(x^2-B)^{\frac{1}{2}}}{(k-x)^{\frac{3}{2}}}
(k-x-4\delta_{3 l'}^2)^{\frac{1}{2}} \left[(k-x-4\delta_{3 l'}^2)x(k-x) \right.
\nonumber \\
 & + & \left. (k-x+2\delta_{3 l'}^2)(2kx-x^2-B) \right]dx \label{r33}
\end{eqnarray}
with

\begin{equation}
k=1+\delta^2_{ll'},\quad B=4(k-1),\quad
\delta_{3l'}=\frac{m_{\nu_3}}{m_{l'}},\quad
\delta_{ll'}=\frac{m_l}{m_{l'}},
\label{def1}
\end{equation}

\begin{equation}
x_m=2\delta_{ll'},\quad
x_M=k-\frac{(m_{\nu_i}+m_{\nu_j})^2}{m^2_{l'}}.
\label{def2}
\end{equation}
$\Gamma^{l'l}_{00}$, $\Gamma^{l'l}_{03}$ and $\Gamma^{l'l}_{33}$
are the contributions for the $l'\to l\bar\nu_i\nu_j$ decays
from two massless, one massive and two massive neutrinos,
respectively.

Using explicitly the parameterization in (\ref{ckm}) we obtain
\begin{eqnarray}
\Gamma(\mu\to e \nu_\mu\bar\nu_e)&=&\frac{G^2m^5_{\mu}}{192\pi^3}\left[
(s_\beta^2s_\gamma^2+c_\gamma^2)c_\beta^2\,\Gamma^{\mu e}_{00}
+(s_\beta^4s_\gamma^2+
s_\beta^2c_\gamma^2+c_\beta^4s_\gamma^2)\Gamma^{\mu e}_{03}\right.\nonumber \\
&+&
\left.s_\gamma^2c_\beta^2s_\beta^2\,\Gamma^{\mu e}_{33}
\right]
\label{mue}
\end{eqnarray}

\begin{eqnarray}
\Gamma(\tau\to e
\nu_\tau\bar\nu_e)&=&\frac{G^2m^5_{\tau}}{192\pi^3}\left[(
s_\beta^2c_\gamma^2+s_\gamma^2)c_\beta^2\,\Gamma^{\tau e}_{00}
+(s_\beta^4c_\gamma^2
+s_\beta^2s_\gamma^2+
c_\beta^4c_\gamma^2)\,\Gamma^{\tau e}_{03}\right.\nonumber \\ &+&\left.
c_\gamma^2c_\beta^2s_\beta^2\,\Gamma^{\tau e}_{33}
\right],
\label{taue}
\end{eqnarray}

\begin{eqnarray}
\Gamma(\tau\to \mu
\nu_\tau\bar\nu_\mu)&=&\frac{G^2m^5_{\tau}}{192\pi^3}\left[
(s_\beta^2c_\gamma^2+s_\gamma^2)(s_\beta^2s_\gamma^2+c_\gamma^2)\,
\Gamma^{\tau\mu}_{00}
\right.\nonumber
\\ &+&\left. \left(s_\gamma^2(s_\beta^2c_\gamma^2+
s_\gamma^2)+c_\gamma^2(s_\beta^2s_\gamma^2+c_\gamma^2) \right)c_\beta^2\,
\Gamma^{\tau\mu}_{03}
+c_\gamma^2s_\gamma^2 c_\beta^4\,\Gamma^{\tau\mu}_{33}\right]
\label{taumu}
\end{eqnarray}
Notice that Eqs.~(\ref{mue})-(\ref{taumu}) depend only on the angles
$\beta$ and $\gamma$. We see from Eq.~(\ref{mue}) that in fact
\begin{equation}
G^2_\mu=G^2\times factor
\label{gmu}
\end{equation}
where the $factor$ will depend on the kinematical region allowed for
a particular decay, so
that the constant $G^2$ appearing in other decays is that given by
Eq.~(\ref{gmu}). For instance, if $\nu_3$ is kinematically forbidden
in the muon decay, from (\ref{mue}) we obtain
\begin{eqnarray}
G_\mu^2&=&G^2(1-\vert V_{\mu3}\vert^2)(1-\vert V_{e3}\vert^2)
\nonumber \\ &= &G^2(s_\beta^2s_\gamma^2+c_\gamma^2)c_\beta^2.
\label{gmu2}
\end{eqnarray}
Hence, we have defined $G_\mu$ in such a way that it
coincides with the usual value.

We will write Eqs.~(\ref{mue}), (\ref{taue}) and (\ref{taumu}),
 respectively, in the following form
\begin{equation}
\Gamma(\mu\to e\nu_\mu\bar\nu_e)=\frac{G^2m^5_\mu}{192\pi^3}f^{\mu
e}(\beta,\gamma,\delta_{e\mu},\delta_{3\mu}),
\label{def3}
\end{equation}
for the partial rate of the muon decay into electron and
\begin{equation}
\Gamma(\tau\to e\nu_\tau\bar\nu_e)=\frac{G^2m^5_\tau}{192\pi^3}f^{\tau
e}(\beta,\gamma,\delta_{e\tau},\delta_{3\tau}),
\label{def4}
\end{equation}

\begin{equation}
\Gamma(\tau\to \mu\nu_\tau\bar\nu_\mu)=\frac{G^2m^5_\tau}{192\pi^3}f^{\tau
\mu}(\beta,\gamma,\delta_{\mu\tau},\delta_{3\tau}),
\label{def41}
\end{equation}
for the decay of $\tau$ into electron and muon, respectively.

Using Eqs.~(\ref{def3})-- (\ref{def41}) we obtain
\begin{equation}
\left(\frac{G_\tau}{G_\mu}\right)^2 = \frac{f^{\tau
e}(\beta,\gamma,\delta_{e\tau},\delta_{3\tau})}{f^{\mu
e}(\beta,\gamma,\delta_{e\mu},\delta_{3\mu})}.
\label{tu}
\end{equation}
and

\begin{equation}
\frac{B^\tau_\mu}{B^\tau_e}= \frac{f^{\tau
\mu}(\beta,\gamma,\delta_{\mu\tau},\delta_{3\tau})}{f^{\tau e}
(\beta,\gamma,\delta_{e\tau},\delta_{3\tau})}=\frac{\Gamma^\tau_\mu}
{\Gamma^\tau_e}.
\label{taufra}
\end{equation}
The left hand sides of (\ref{tu}) and (\ref{taufra}) have the value showed
in (\ref{tmu})  and (\ref{tmu2}) respectively.

We will also consider $\pi\to l+\nu_l,\;l=e,\mu$ decays. The partial
width with a massive neutrino is~\cite{ro}
\begin{equation}
\Gamma^{\pi l}_3=\frac{G^2f^2_\pi V^2_{KM} m^3_\pi}{8\pi}
\left[\delta^2_{l\pi}+\delta^2_{3l}-(\delta^2_{l\pi}-\delta^2_{3l})^2\right]
\lambda^{\frac{1}{2}}(1,\delta^2_{l\pi},\delta^2_{3l})
\label{piw}
\end{equation}
where $\lambda$ is the triangular function,
$\delta_{l\pi}=m_l/m_\pi$ and $V^2_{KM}$ is the appropriate
Cabibbo-Kobayashi-Maskawa matrix elements of the quark sector.
In particular we have
\begin{eqnarray}
\Gamma(\pi\to \mu\nu_\mu)&=&\left(\vert
V_{\mu1}\vert^2+ \vert V_{\mu2}\vert^2 \right)\Gamma^{\pi\mu}_0+
\vert V_{\mu3}\vert^2\Gamma^{\pi\mu}_3\nonumber \\&=&
\left(s_\gamma^2s_\beta^2+c_\gamma^2 \right)\Gamma^{\pi\mu}_0+
s_\gamma^2c_\beta^2
\Gamma^{\pi\mu}_3
\label{pimu}
\end{eqnarray}
\begin{eqnarray}
\Gamma(\pi\to e\nu_e)&=&\left( \vert
V_{e1}\vert^2 +\vert V_{e2}\vert^2\right)\Gamma^{\pi e}_0+
\vert V_{e3}\vert^2 \Gamma^{\pi e}_3\nonumber \\ &=&
c_\beta^2\Gamma^{\pi e}_0+ s_\beta^2 \Gamma^{\pi e}_3
\label{pie}
\end{eqnarray}
{}From (\ref{pimu}) and (\ref{pie}) we obtain
\begin{equation}
\frac{\Gamma^{\pi e}}{\Gamma^{\pi\mu}}=\frac{c_\beta^2\Gamma^{\pi e}_0+
s_{\beta}^2 \Gamma^{\pi e}_{3}}
{(s_\gamma^2s_\beta^2+c_\gamma^2)\Gamma^{\pi\mu}_0+s_{\gamma}^2
c_{\beta}^2 \Gamma^{\pi \mu}_3}=
\frac{B^{\pi e}}{B^{\pi\mu}},
\label{piemu}
\end{equation}

In the next section we will compare Eqs.~(\ref{tu}), (\ref{taufra})
and (\ref{piemu}) with (\ref{tmu})-(\ref{pi}) and other experimental
data.
\section{Experimental constraints on mixing angles and neutrino mass}
\label{sec:ma}

The effects of a leptonic--mixing
in the decays of pseudoscalars mesons was studied by Shrock~\cite{ro}
and Ng~\cite{ng}. Since then, several experiments were performed
searching the effect of extra peaks due to massive neutrinos in the
spectrum of the positron in $\pi^+\to e^+\nu(\gamma)$~\cite{b1}. The
measurement of this branching ratio
confirm the hypothesis of the $e-\mu$ universality at the $0.2\%$ level
{}~\cite{b2}. The theoretical uncertainty arises from
structure--dependent loop effects, they could be important in the
$\pi\to e\nu_e\gamma$ decay, but are less than $0.1\,\%$~\cite{b2}.

Here we will point out
some considerations about these experiments. In most of them
when fitting experimental data with theoretical calculations
it have been assumed that the heavy
neutrino couples mainly with the
electron or muon. Hence, according to this what they are comparing
with experimental data is~\cite{dib2}
\begin{equation}
\frac{\Gamma(\pi\to l\nu_3)}{\Gamma(\pi\to l\nu_l)}\,
\propto \vert V_{l3}\vert^2
\label{expe}
\end{equation}
where $\nu_l$ is the conventional massless neutrino, $l=e,\mu$ and the
factor of proportionality is the kinematical factor including phase
space.

It has also been consider by Bryman et al. ~\cite{b1}, the  ratio
\begin{equation}
\frac{\Gamma(\pi\to e\nu_3)}{\Gamma(\pi\to \mu\nu_l)}
\label{}
\end{equation}
which depends on both $V_{e3},V_{\mu3}$. However, the value of $V_{\mu3}$
was taken from the work by Abela et al.~\cite{b1}. In fact, most of the
experimental data were fitted assuming a heavy
subdominant coupled (HSC) neutrino, i.e., a heavy neutrino with
$V_{li}\ll1$~\cite{nota1}.

In our analysis we will not make any assumption with
respect to the matrix elements of the mixing matrix and when we have
considered a kinematically allowed heavy neutrino in pion decays
it is necessary to use (\ref{pimu}) and (\ref{pie}) in which
$V_{e3,\mu3}$ appear (after using the unitarity condition on the
matrix $V$).

We have done several plots of Eqs.~(\ref{tu}), (\ref{taufra})
and (\ref{piemu}) as a
function of the angles $\beta$ and $\gamma$ for several values of
$m_{\nu_3}$ using the experimental value (within $1\sigma$) given by
(\ref{tmu}), (\ref{tmu2}) and (\ref{pi}). The values of the other
parameters are those of the Ref.~\cite{pdg}.

Let us consider the following intervals :

\noindent {\em i)} $G_\tau/G_\mu>1$ independently of the mixing angles.
For the experimental data we have used, this region corresponds to $m_{\nu_3}
<50$ MeV.

\noindent {\em ii)} $G_\tau/G_\mu\leq1$, but there is no
intersection with (\ref{tmu}). This corresponds to the
mass interval $50\,\mbox{MeV}<m_{\nu_3}<84\,\mbox{MeV}$.

\noindent {\em iii)} $G_\tau/G_\mu\leq1$, compatible with (\ref{tmu})
and (\ref{tmu2}) but not compatible with (\ref{pi}). In this case we have
$84\,\mbox{MeV}\leq m_{\nu_3}<155\,\mbox{MeV}$ and $m_{\nu_3}>800$ MeV.

\noindent {\em iv)} $G_\tau/G_\mu\leq1$, compatible with all
experimental data in the $m_{\nu_3}$ interval: $155\,\mbox{MeV}\leq
m_{\nu_3}\leq800\,\mbox{MeV}$.

In Fig. 1 we show contour plot  of (\ref{tu}),
(\ref{taufra}) and
(\ref{piemu}). The allowed values for the mixing angles we obtain,
using $m_{\nu_3}=165\,\mbox{MeV}$, are
\begin{equation}
11.54^\circ<\beta<12.82^\circ,\quad \gamma<4.05^\circ.
\label{angles}
\end{equation}
We have seen that, to be in agreement with the processes represented by the
quantities in Eqs.~(\ref{tu}), (\ref{taufra})
and (\ref{piemu}) the heavier neutrino could have a mass in the
interval:
\begin{equation}
155 \,\mbox{MeV}\lower.7ex\hbox{$\;\stackrel{\textstyle<}{\sim}\;$}
m_{\nu_3} \lower.7ex\hbox{$\;\stackrel{\textstyle<}{\sim}\;$}800\,
\mbox{ MeV}.
\label{m3}
\end{equation}

\section{Constraints coming from $Z$- invisible width and end point of
tau decay into five pions}
\label{zotau}
We have seen in the last section that the mass region allowed by
the processes (\ref{tmu}), (\ref{tmu2}) and (\ref{pi}) is given in
(\ref{m3}). However, it is  mandatory to consider the $Z^0$ invisible
width and also the end point of the spectrum of the $\tau\to
5\pi\nu_\tau$ decay which provide the current upper limit on the
``tau neutrino'' mass. Hence, we will continue
our analysis in order to constrain even
more the allowed values for the neutrino mass and mixing angles.
\subsection{$Z^0$ invisible width}
\label{subsec:z0}
Let us consider the constraints coming from the $Z^0$ invisible width
measured by the LEP experiments: $\Gamma^{inv}=502\pm9$ MeV~\cite{pdg}.
For two massless and one massive neutrinos we have~\cite{hollik}
\begin{equation}
\frac{\Gamma^{inv}}{\Gamma^0} =
\left[2+\sqrt{1-4\mu_F}\,(1-\mu_F)\right]
\frac{1}{[(s_\beta^2s_\gamma^2+c_\gamma^2)c_\beta^2]^{\frac{1}{2}}},
\label{width}
\end{equation}
where
$$\Gamma^0= \frac{M_Z^3 G_\mu}{12\sqrt{2}\pi},\quad
\mu_F=m^2_{\nu_3}/M^2_Z,$$
$$
M_Z = 91.173\pm 0.020 \, \mbox{GeV} \, \mbox{
and} \, G_\mu =
1.16639\times 10^{-5}\,\mbox{GeV}^{-2}.$$
and we have used (\ref{gmu2}).

The contour plot of Eq.(\ref{width}) for $m_{\nu_3}=165 $ MeV appears also
in Fig. 1.
For masses above 225 MeV there is no compatibility region among $Z^0$
data and the other experimental ratios we have considered
in Sec.~\ref{sec:ma}. Hence, instead of (\ref{m3}) we have now
\begin{equation}
155\,\mbox{MeV}\lower.7ex\hbox{$\;\stackrel{\textstyle<}{\sim}\;$} m_{\nu_3}
\lower.7ex\hbox{$\;\stackrel{\textstyle<}{\sim}\;$}225\, \mbox{ MeV}.
\label{m3b}
\end{equation}
For now on in our analysis we will use a typical value of $m_{\nu_3}
=165$ MeV unless otherwise stated.

\subsection{End point of $\tau\to 5\pi\nu_\tau$ in the mixing scenario}
\label{subsec:bound}
The current upper limits on neutrino masses are:
\begin{equation}
\begin{array}{c}
``m_{\nu_e}''<7.3\,\mbox{eV}\,\quad \cite{pdg},\\
``m_{\nu_\mu}''<0.27\,\mbox{MeV}\, \quad \cite{pdg},\\
``m_{\nu_\tau}''<31\,\mbox{MeV}\, \quad \cite{argus,argus2}.
\end{array}
\label{numass}
\end{equation}
However these values were obtained in experiments which
tried to observe the effects of neutrino masses without
mixing in the leptonic sector.

The current experimental upper bound on ``$m_{\nu_\tau}$'' given by
ARGUS~\cite{argus,argus2} come from the study of the end point of the
hadronic invariant mass
distribution in the decay $\tau\to 5\pi\nu_\tau$.
Their analysis is only valid when there is no mixing in the leptonic sector,
otherwise the massive neutrino will not manifest itself by a shift on the
end point but by a shoulder in the distribution. The end point in both
situation
   s
will be the same.

For the sake of simplicity we shall consider here only the $\tau$
neutrino case but a similar procedure could be applied to the other $e$
and $\mu$ neutrinos. Assuming that $\nu_\tau$ in the decay $\tau\to
n\pi\nu_\tau$ is a particle with definite mass it was
obtained~\cite{concha}
\begin{equation}
\frac{d\Gamma}{dq^2}=\frac{G^2V^2_{KM}}{8\pi m_\tau^3}
w(q^2,m_\tau^2,m_{\nu_\tau}^2)\lambda^{\frac{1}{2}}
(m_\tau^2,q^2,m_{\nu_\tau}^2)h(q^2),
\label{gamma1}
\end{equation}
where
\begin{equation}
w(q^2,m_\tau^2,m_{\nu_\tau}^2)=(m_\tau^2-q^2)(m_\tau^2+2q^2)-
m_{\nu_\tau}^2(2m_\tau^2-q^2-m_{\nu_\tau}^2),
\label{w}
\end{equation}
and
\begin{eqnarray}
\lambda^{\frac{1}{2}}(m_\tau^2,q^2,m_{\nu_\tau}^2)&=&\left[ \right.
\{m_\tau^2-[(q^2)^{\frac{1}{2}}+m_{\nu_\tau}]^2\}\nonumber \\ &
&\mbox{} \times\{m_\tau^2-[(q^2)^{\frac{1}{2}}-m_{\nu_\tau}]^2\}
\left.\right]^{\frac{1}{2}}.
\label{l}
\end{eqnarray}
In Eq.~(\ref{gamma1}) the
function $h(q^2)$ contains the hadronic structure and $V^2_{KM}$
denotes the quark mixing angles, in this case $V^2_{KM}=V^2_{ud}$.
Here we shall not write down
explicitly both factors~\cite{concha}.

At the end point of the
spectrum
\begin{equation}
m_{\nu_\tau}=m_\tau-m_{had},
\label{e}
\end{equation}
where $m_{had}=(q^2)^{\frac{1}{2}}$.
Assuming that the $\nu_3$ is heavier than $\nu_{1,2}$ we get, instead
of Eq.(\ref{gamma1})
\begin{equation}
\frac{d\Gamma}{dq^2}=\frac{G^2V^2_{KM}}{8\pi m_\tau^3}\left[ (\vert
V_{\tau1}\vert^2+\vert
V_{\tau2}\vert^2) F_0 +\vert
V_{\tau3}\vert^2 F_3
\right]h(q^2)
\label{gamma2}
\end{equation}
where
\begin{equation}
F_0=  w(q^2,m_\tau^2,0)\lambda^{\frac{1}{2}}
(m_\tau^2,q^2,0)
\label{0}
\end{equation}
and
\begin{equation}
F_3=w(q^2,m_\tau^2,m_{\nu_3}^2)
\lambda^{\frac{1}{2}}(m_\tau^2,q^2,m_{\nu_3}^2).
\label{3}
\end{equation}
Each part of the phase space functions $F_0,F_3$ have a different end point.

The second end point related to the massive contribution $m_{\nu_3}$
disappears when both massive and massless contributions are summed up,
remaining only a shoulder in the distribution and the end point is related
with the (almost) massless neutrinos. This is the end point observed
experimentally.

Using $m_{\nu_3}=165$, $225$ MeV and typical
values for the respective mixing angles in Eq.(\ref{gamma2}) we
obtain Fig. 2. We see that the $m_{\nu_3}$ values given in (\ref{m3b})
are allowed by the experimental data of the $\tau\to5\pi\nu$ decay.
We also show in Fig. 2 a blow-up of the region near the end point to emphasize
the fact that all the three curves for $m_{\nu_3}=0,165,225$ MeV would be
compatible with the three events found by ARGUS in this end point
area~\cite{argus,argus2}.

We have only studied the phase space contributions and have not taken
into account the hadronic effects contained in $h(q^2)$. This
function hides our ignorance on hadronic structure and it will, in
general, modify the shape of the distribution but not affect the
end point. However close to the end point the shape of the distribution
does not depend very much on the resonant structure. On the other
hand, the population of the spectrum in this region will be strongly
modified~\cite{concha}. For this reason, to be able to detect an
effect of mixing it is necessary to study the population of the
spectrum close to the end point.

\section{Other constraints}


\subsection{neutrino oscillations}
\label{subsec:osc}
The case of three light neutrino oscillations was consider in
Ref.~\cite{vbarger}. Here we will study the case of only two light
neutrinos and a heavy one as in Ref.~\cite{ng}. Let us start by
summarizing at this stage our results for the mixing angles:
\begin{equation}
\left(
\begin{array}{ccc}
0.977\,c_\theta & 0.977\, s_\theta  &  0.211 \\
-0.998\,s_\theta-0.016\,c_\theta  & 0.998\,c_\theta-0.016\,s_\theta  & 0.069
\\
0.071\,s_\theta-0.2\,c_\theta  & -0.071\,c_\theta-0.211\,s_\theta & 0.975
\end{array}\right).
\label{ckmn}
\end{equation}
The matrix above is orthogonal independently of the $\theta$.

As we said before, neutrinos produced in weak processes are linear combinations
of the mass eigenstates $\nu_l=\sum_iV_{li}\nu_i$ with $V_{li}$ given in
(\ref{ckm}) and in particular in (\ref{ckmn}). Here we will write
down the probability of finding a neutrino $\nu_e,\nu_\mu$ after a
path length $R$ if at the origin it was a $\nu_e$, that is
\begin{eqnarray}
P(\nu_e\to\nu_e)=
\vert\langle\nu_e(L)\vert\nu_e(0)\rangle\vert^2&=&
1-2c_\beta^2s_\beta^2\left(1-\cos\frac{2\pi R}{L}\right)\nonumber \\
&-&2c_\theta^2s_\theta^2c_\beta^4\left(1-\cos\frac{2\pi R}{L_{12}}\right),
\label{ee}
\end{eqnarray}
and
\begin{eqnarray}
P(\nu_e\to\nu_\mu)&=&
\vert\langle\nu_\mu(L)\vert\nu_e(0)\rangle\vert^2 \nonumber \\&=&
h_1(\beta,\gamma,\theta)-
h_2(\beta,\gamma,\theta)\cos\frac{2\pi
R}{L_{12}}-h_3(\beta,\gamma,\theta)\cos\frac{2\pi R}{L}
\label{emu}
\end{eqnarray}
where we have defined
\[
h_1(\beta,\gamma,\theta)=2c_\beta^2[c_\theta^2s_\theta^2c_\gamma^2+
(1-c_\theta^2s_\theta^2)s_\gamma^2s_\beta^2+
c_\gamma^2s_\gamma^2s_\beta^2c_\theta^2s_\theta^2(c_\theta^2-s_\theta^2)],
 \]
\[
h_2(\beta,\gamma,\theta)=2[c_\theta s_\theta(c_\gamma^2-s_\gamma^2s_\beta^2)
+c_\gamma s_\gamma s_\beta(c_\theta^2-s_\theta^2)]c_\theta s_\theta c_\beta^2,
\]
\[
h_3(\beta,\gamma,\theta)=2s_\gamma^2s_\beta^2c_\beta^2.
\]

As in Ref.~\cite{ng}, for the mass range we are considering here,
neutrino oscillations occur with essentially two wavelengths. We have
defined in (\ref{ee}) and (\ref{emu})
\[ L_{ij}=\frac{2\pi}{E_i-E_j}\]
that is,
\[L_{12}=\frac{4\pi p}{\delta m^2}=2.5\frac{p/\mbox{MeV}}{\delta
m^2/(\mbox{eV})^2}\,\mbox{meter}, \]
with
\[\delta m^2=m^2_{\nu_1}-m^2_{\nu_2},\]
and
\[
L_{13}=L_{23}=L=\frac{2\pi}{m_{\nu_3}}=\frac{1.24}{m_{\nu_3}/\mbox{MeV}}
\times10^{-12}\,\mbox{meter}.\]
The masses
$m_{\nu_1}, m_{\nu_2}$, and of course $\delta m^2$, are still
undetermined. Let us take,for instance, $\delta m^2$ of the order of a few
$\mbox{eV}^2$. Of course for smaller $\delta m^2$ values the cosine involving
$L_{12}$  in Eq.(~\ref{ee}) must be taken into account.
The short
component has an oscillation length of the order of $10^{-15}$ m.
  A more detailed study
concerning neutrino oscillation data will be published elsewhere.

Let us consider the solar neutrino data. In this case
$R\sim10^{11}$ m and $L_{12}\approx 0.25$ m if $\delta m^2\approx\,
1(\mbox{eV})^2$. With this condition the oscillation has averaged
out and one obtain,
\begin{equation}
P(\nu_e\to\nu_e)=1-2c_\beta^2s_\beta^2-2c_\theta^2s_\theta^2c_\beta^4.
\label{solarav}
\end{equation}
Using the values in (\ref{ckmn}) and the experimental data given in Table 1
for $P(\nu_e\to\nu_e)$~\cite{osc} we
obtain the allowed regions showed in Fig. 3.
In Fig. 4 we show values for the angles compatible with those in
Fig. 1.
Note that with
these angles the result of Davis, a suppression of about $27\%$ of the
solar standard model prediction,
is not fitted at the $1\sigma$ level.
The lowest value of Eq.(\ref{solarav}) is about $ 30\%$.

In
fact, neutrino oscillations are only sensitive to small mass
differences and cannot be used to detect mixing of heavy neutrinos
directly. However, we see from (\ref{solarav}) that even in the
averaged situation the effect of the mixing with the heavy neutrino
survives via the angle $\beta$.

The range for the $\theta$ angle compatible with the allowed area
showed in Fig.1, coming from Kamiokande II, Gallex and SAGE data is
\begin{eqnarray}
 & 0.25271<s_{\theta}^2< 0.29862, \nonumber \\
 & 0.70142<s_{\theta}^2<0.74729.
\label{kamioka}
\end{eqnarray}

Finally, the mixing matrix in the leptonic
sector which is consistent with all experimental data considered
by us, for $m_{\nu_3}=165$ MeV, is:
\begin{equation}
\left(
\begin{array}{ccc}
0.817-0.848\, &\, 0.490-0.536 \, & \, 0.200-0.222 \\
-(0.502-0.560)\,  &\, 0.828-0.865\,  & \,<0.069 \\
-(0.192-0.129)\,  &\, -(0.101-0.183)\, & \,0.973-0.980
\end{array}\right).
\label{kmf}
\end{equation}
\subsection{Cosmological constraints}
\label{subsec:cosmo}
{}From cosmological constraints, stable neutrinos must have masses less
than 40 eV or greater than 2 GeV~\cite{lw}. In our scenario, of course,
by construction neutrinos are not stable.

Earlier works about the
upper bounds on the lifetime of a massive neutrino assumed V-A
interactions in annihilation processes of the massive neutrinos and
that the principal decay mode is $\nu_3\to
\nu_e\gamma$~\cite{dicus,sarkar}. However, for a neutrino mass
$m_{\nu_3}\gg m_{e,\mu}$ this decay (and the not GIM suppressed $\nu_3\to
\nu_e\gamma\gamma$) are not dominant~\cite{sarkar}.

Note that as there are no changing flavor neutral currents at tree level,
constraints as those in Ref.~\cite{aan} do not apply to our case.

Here we will assume the mixing angles given in (\ref{kmf}) and
evaluate the lifetime for a heavy neutrino.
In the mass range we are considering for the heaviest neutrino,
$155-225$ MeV, the main decays are $\nu_3\to l^+l^-\nu_{1,2}$,
$l=e,\mu$. The lifetime of heavy neutrinos in our context
was studied by Kolb and Goldman~\cite{kg}. Here we
will treat again these decays. The partial widths are
\begin{equation}
\Gamma(\nu_3\to ll\nu)=\frac{G_\mu^2 m^5_{\nu_3}}{192\pi^3
(1-\vert V_{\mu3}\vert^2)(1-\vert V_{e3}\vert^2)}[\vert
V_{l3}\vert^2 (1-\vert V_{l3}\vert^2)F(m_l,m_{\nu_3})],
\label{wnu3e}
\end{equation}
And a similar expression for $\nu_3\to\mu e\nu$.

Notice that we have taken into account the appropriate modification of
the $G_\mu$ constant given in Eq.~(\ref{gmu2}). In fact, we leave in
(\ref{wnu3e}) all factors in such a way that it is clear that
some cancellation of matrix elements occurs, otherwise we would obtained
only the
factor $\vert
V_{e3}\vert^2 (1-\vert V_{e3}\vert^2)$ for the case of electron decay
 and $\vert
V_{\mu3}\vert^2(1-\vert V_{\mu3}\vert^2)$ for $\mu$ decay as in
ref.~\cite{kg}.

The function $F(m_l,m_{\nu_3})$
is defined as
\begin{eqnarray}
F(m_l,m_{\nu_3})& = &
2\int^{y_M}_{y_m}\frac{(y^2-B')}{(k'-y)^{3}}
(k'-y-\omega^2_l)^2 \left[(k'-y-\omega_l^2)y(k'-y) \right.
\nonumber \\
 & + & \left.\left[(k'-y)^2+\omega_l^2(k'-y)-2\omega_l^4\right](2k'y-y^2-B')
\right] dy \label{l33}
\end{eqnarray}
where
\[k'=1+\omega^2_l,\quad B'=4(k'-1),\quad
\omega_l=\frac{m_l}{m_{\nu_3}},\quad y_m=2\omega_l,\quad y_M=1.
\]
Here we will consider only the $\nu_3 \to ee \nu$ decay.
The respective lifetime scaled from muon decay is
\begin{eqnarray}
\tau(\nu_3\to ee\nu)&=&\tau_\mu\left(\frac{m_\mu}{m_{\nu_3}}\right)^5
\frac{1-\vert V_{\mu3}\vert^2}{\vert V_{e3}\vert^2}
\frac{\Gamma^{\mu e}_{0 0}}{F(m_e,m_{\nu_3})}\nonumber \\ &=&
\tau_\mu\left(\frac{m_\mu}{m_{\nu_3}}\right)^5
\frac{1-s_\gamma^2c_\beta^2}{s_\beta^2}
\frac{\Gamma^{\mu e}_{0 0}}{F(m_e,m_{\nu_3})},
\label{vidae}
\end{eqnarray}
where $\tau_\mu$ is the muon lifetime. From (\ref{vidae})
 and the angles in (\ref{kmf}) we see that the lifetime for the decay of
$\nu_3$ into electrons is of the order of $10^{-6}-10^{-5}$ s.

For the range (4.2) the decays
$\nu_3\to\pi^+e^-$ and $\nu_3\to\mu^+
e^-\nu$ are possible, they will not be considered here. Our purpose is only to
indicate that there is no strong constraints coming
from processes with cosmological consequences.

There are experiments which tried to observe neutrino decays~\cite{fb,bd},
however those experiments are sensitive to lifetimes of the order of
$10^{-2}-10^{2}$ s. Hence, their data are not directly
applicable to our case.

For instance,
in \cite{fb} it was assumed that the $\nu_\tau$ couples mainly to a
single mass eigenstate, then an upper limit of the square of the
matrix element $\vert V_{ei}\vert$ was obtained. In our case it is
necessary to consider at least two of such matrix elements at once.

On
the other hand, we must recall the following. In \cite{fb} the
expected neutrino flux
depends on the number of $D_s$ ($N_{D_s}$) and
of $D$ mesons ($N_D$)
produced by protons in the dump. In computing this number it was
assumed the following values for the branching ratio for the semileptonic
decays $B.R.(D\to\mu\nu_3)=0.1$ and $B.R.(D_s\to \mu\nu_3)=0.03$.
Notwithstanding, at present these branching ratios are
$B.R.(D\to\mu\nu_3)\leq7.2\times10^{-4}$~\cite{pdg} and
$B.R.(D_s\to\mu\nu_3)=(4.0^{+1.8+.8}_{-1.4-.6}\pm1.8)\times10^{-3}$~
\cite{refx}. It means that the neutrino flux is reduced by
a factor of a hundred with respect to the values used by the CHARM
experiment~\cite{fb}, and that in fact higher values for the upper bounds
on the mixing angles are allowed.
On the other hand, in Ref.~\cite{bd} the neutrino flux
was calculated using Monte Carlo
programs and we do not know if in this case the same branching ratios
for the decays of the $D$ mesons were used.

Anyway, we stress that in our case the lifetime of the heavy
neutrino is such that it would decay before the detector.

\subsection{astrophysical constraints}
\label{subsec:astro}
Here we will discuss, briefly, the possible effects of our scenario in
the solar neutrino flux and in the supernova 1987A data. In fact for neutrinos
with mass $1.1\,\mbox{MeV}<m<14\,\mbox{MeV}$ the experimental
detection of solar neutrinos have been considered in Ref.~\cite{dtfw}.
Neutrinos coming from the sun, generated through the decay $^8B\to
^8\!B_ee^+\nu$ must have a mass up to near 14 MeV which is the threshold
energy for this reaction. Those coming from $pp$ reaction have
$m<1.44$ MeV. Of
course, a heavy neutrino would decay in flight producing
$\nu_{e,\mu}$ and charged leptons.

The observation of $\bar\nu_e$'s coming from the supernova 1987A
implies a constraint on their lifetime
\begin{equation}
\tau_{\nu_e}/m_{\nu_e}\lower.7ex\hbox{$\;\stackrel{\textstyle>}{\sim}\;$}
6\times10^5s/eV.
\label{87a}
\end{equation}
However this constraint is valid for neutrinos reaching the Earth's
orbit, and no mixing in the
leptonic sector. In fact, we have seen in Sec.~\ref{subsec:cosmo} that for a
mass of 165 MeV the mixing angles are such that the lifetime of the
heavy neutrino is of the order of $10^{-6}-10^{-5}$ s. It means that
this neutrino can appear only through the mixing effect as it is
kinematically forbidden at the characteristic energies of the
supernova and must decay before reaching the Earth's orbit.
We see that only the lightest neutrinos $\nu_{1,2}$ are
constrained from these data.

On the other hand, since neutrinos are massive Dirac particles the
right-handed components can be produced in the hot supernova via
helicity flip. This produce an extra source of cooling in the
supernova since this right-handed neutrinos are almost sterile.
The relevant spin-flip processes are $NN\to NN\bar\nu_R\nu_R$
(bremsstrahlung), $N\nu_L\to N\nu_R$ and $e^+e^-\to\bar\nu_R\nu_R$.
However bounds coming from this  processes in supernova are valid
only for small enough neutrino masses and do not apply to our
heavy neutrino. It is not thermally emitted from
the supernova~\cite{gri,ra}.

\section{conclusions}
\label{sec:con}
We have
examined some weak processes which are modified by a three generation
mixing in the leptonic sector and
found compatibility regions for $m_{\nu_3}$ and mixing angles.
These parameters are likely to have consequences elsewhere.

The main result of this paper is that the current upper limit on the
tau-neutrino mass ``$m_{\nu_\tau}$'' does not seem to exclude a
heavy neutrino $155\,\,\mbox{MeV}<m_{\nu_3}
<225\,\,\mbox{MeV}$ mixed up
with the light ones, and solving at the same time a possible deviation of the
$\tau-\mu$ universality. However, our analysis
will still be valid if no deviation from $\tau-\mu$ universality is confirmed
by data. Only the value for $m_{\nu_3}$ and mixing angles will be different.
If the new data imply lowest values for the $\nu_3$ mass
and, as in $\nu_3$ decays photons are ultimately
produced~\cite{dolgov} it will be interesting to study
astrophysical bounds more in detail.

Of course, there are other experiments we could have considered as the
$\nu_\mu e\to\nu_\mu e$ cross section ~\cite{mg1}, measurements of
the Michel parameter in muon decay ~\cite{msd}, etc., but we do not
expect they to produce strongest constraints than the ones we have analyzed.

The $\mu-e$ universality has been also verified in the kaon
decays~\cite{kaons1}. In fact, in the mass range (\ref{m3b})
the decays $K\to e(\mu)\nu_i$ with
$i=1,2,3$ occur.
In particular, the branching
ratio of the kaon decays, $\Gamma(K\to e\nu)/\Gamma(K\to \mu\nu)$
have been measured~\cite{kaons2} and, in principle, provide another
confirmation of the $e\mu$ universality, but in this case the
structure--dependent radiative correction is expected to be
$\sim1000$ times larger than that for the pion decays partial widths
{}~\cite{b2}. For this reason we will not analyze these decays
in this work.
However we recall  that in the literature a similar expression to
(\ref{expe}) was also assumed.

Notice that in the extension of the standard model we are considering
there are no flavor changing neutral currents at tree level, since
they are GIM suppressed~\cite{gim}, and for this reason it is not necessary
to consider processes producing $\mu e$ events.

We recall that the final allowed values for masses
and mixing angles must be obtained taken into account radiative
corrections~\cite{pk2}.
Any way, radiative corrections
imply a reduction of about $\sim4\,\%$~\cite{b2} and will
not modify qualitatively our results.


\acknowledgments
We thank Funda\c c\~ao de Amparo \`a Pesquisa do Estado de
S\~ao Paulo (FAPESP) (O.L.G.P.) for full financial support and
Con\-se\-lho Na\-cio\-nal de De\-sen\-vol\-vi\-men\-to Cien\-t\'\i
\-fi\-co e Tec\-no\-l\'o\-gi\-co (CNPq) (V.P.) for partial and full (R.Z.F.)
financial support. We are very grateful to C.O. Escobar for stimulating
discussi
   ons
at the early stage of this work. We also thank M.M. Guzzo for very useful
comments concerning solar neutrino data.

\begin{figure}
\setepsfscale{0.78}
\begin{picture}(700,400)(0,0)
\postscript {fig1.ps}
\end{picture}
\protect
\caption[Fig.1]{Allowed region (within $1\sigma$) in the plane $s^2_\beta
\times s^2_\gamma$ for the ratios $G_\tau/G_\mu$
(Eq.~(\ref{tmu})),$B^{\tau \mu}/B^{\tau e}$ (Eq.~(\ref{tmu2})) and $B^{\pi
e}/B^{\pi\mu}$ (Eq.~(\ref{pi})) for $m_{\nu_3}=165 \mbox{ MeV}$. The region
 for $B^{\tau \mu}/B^{\tau e}$  is larger than the area shown in the
picture. We will also show the contour plot of (Eq.~(\ref{width})).
The black area is the allowed region combining all ratios.}
\label{f1}
\end{figure}
\begin{figure}
\setepsfscale{0.77}
\begin{picture}(700,554)(0,0)
\postscript {fig2.ps}
\end{picture}
\caption[Fig.2]{Spectrum for $m_{\nu_3}=0$ (dashed line), $m_{\nu_3}=165
\mbox{ MeV}$ (dotted line) and $m_{\nu_3}=215 \mbox{ MeV}$ (continuous line).
Here $d\Gamma^*/dx=(8\pi m_\tau^3/G^2V_{KM})^{-1}d\Gamma/dx$, $G^2=
G^2_\mu/(s_\beta^2s_\gamma^2+c_\gamma^2)c_\beta^2$ and $x=m_{h}/m_{\tau}$.
The coefficient
$(s_\beta^2s_\gamma^2+c_\gamma^2)c_\beta^2$
has been absorbed in the definition of $\Gamma$.
We also shown the blow-up of region near of end point.}
\label{f2}
\end{figure}
\begin{figure}
\setepsfscale{0.87}
\begin{picture}(700,554)(0,0)
\postscript {fig3.ps}
\end{picture}
\caption[Fig.3]{Contour plot of Eq.~(\ref{solarav}) for Sage, Gallex
and Kamiokande--II using the values on Table~\ref{tab1} (within 1 $\sigma$).
 The arrows shown the allowed contours for each experiment.}
\label{f3}
\end{figure}
\begin{figure}
\begin{picture}(700,554)(0,0)
\postscript {fig4.ps}
\end{picture}
\caption[Fig.4]{Contour plot for Gallex, Kamiokande--II and Sage data
in the restricted region showed in black
on Figure~\ref{f1}. The shaded areas are the allowed regions for these
experiments.}
\label{f4}
\end{figure}

\begin{table}[h]
\begin{center}
\begin{tabular}{|l|l|l|l|}
\hline
\bf Experiment & Process & $E_{\mbox{threshold}}$ (MeV) & Expt./SSM \\
\hline
\bf Davis et al. & $\nu_e+^{37}\mbox{Cl} \rightarrow e +^{37}\mbox{Ar}$ & 0.81
&
    $0.27\pm0.04$ \\
\bf Kamiokande--II & $\nu + e \rightarrow \nu + e$ & 7.5 & $0.49\pm0.05\pm0.06$
   \\
\bf SAGE & $\nu_e+^{71}\mbox{Ga} \rightarrow e+^{71}\mbox{Ge}$ &0.24 &
$0.44^{+0
   .13}_{-0.18}\pm0.11$ \\
\bf GALLEX & $\nu_e +^{71}\mbox{Ga} \rightarrow e+^{71}\mbox{Ge}$ &0.24 &
$0.66\
   pm0.11\pm0.05$ \\
\hline
\end{tabular}
\end{center}
\caption[Tab.1]{Results of solar neutrino experiments. The flux is given as a
function of the Solar Standard Model prediction~\cite{fn1}.}
\label{tab1}
\end{table}

\end{document}